\documentclass[12pt]{scrartcl}

\usepackage{fixltx2e}
\usepackage{lmodern}
\usepackage[T1]{fontenc}
\usepackage[utf8]{inputenc}
\usepackage{booktabs}
\usepackage{siunitx}
\usepackage[margin=1.2in]{geometry}
\usepackage{titling}
\usepackage{authblk}
\usepackage{amssymb}
\usepackage{amsmath}
\DeclareMathOperator{\marg}{marg}

\DeclareMathOperator{\chat}{Chat}

\newcommand{\interior}{\text{int}}
\usepackage{graphicx}
\usepackage{amsthm}
\usepackage{array}
\usepackage{hyperref}
\usepackage[dvipsnames]{xcolor}
\usepackage{ascmac}
\usepackage{setspace} 
\usepackage{csquotes}
\usepackage{soul}
\usepackage[authoryear]{natbib} 
\setkomafont{disposition}{\bfseries}

\usepackage{tikz}
\usetikzlibrary{arrows.meta, trees}

\usepackage{sectsty}
\sectionfont{\fontsize{14}{15}\selectfont}
\subsectionfont{\fontsize{13}{15}\selectfont}

\newtheorem{definition}{Definition}
\newcounter{example}
\newenvironment{example}[1][]{\refstepcounter{example}\par\medskip
   \noindent \textbf{Example~\theexample. #1} \rmfamily}{\medskip}

\newtheorem{proposition}{Proposition}
\newtheorem{observation}{Observation}

\newtheorem{lemma}{Lemma}

\theoremstyle{remark}

\newtheoremstyle{named}{}{}{\itshape}{}{\bfseries}{.}{.5em}{\thmnote{#3  }#1}
\theoremstyle{named}

\newcommand{\keywords}[1]{%
  \vspace{1em}
  \noindent\textbf{Keywords:} #1
}

\newcommand{\jelclassification}[1]{%
  \vspace{0.5em}
  \noindent\textbf{JEL Classification:} #1
}

\title{When Truth Does Not Take on Its Shoes: \\ How Misinformation Spreads in Chatrooms\thanks{I am grateful to Pierpaolo Battigalli, Luca Braghieri, Manuel Foerster, Dominik Karos, Lanyi Liu, Fabio Maccheroni, Zsombor Z. M\'{e}der, Isabelle Mortillaro, Christopher Turansick, Yuxin Wang for discussions, advices, and encouragements.}}
\author{Shuige Liu \footnote{Department of Decision Sciences, Bocconi University. Email: \textsf{shuige.liu@unibocconi.it}.}}

\date{\today}

\begin{document}

\maketitle

\begin{abstract}
\textbf{\abstractname.} 
We examine how misinformation spreads in social networks composed of individuals with long-term offline relationships. Especially, we focus on why misinformation persists and diffuses despite being recognized by most as false. In our psychological game theoretical model, each agent who receives a piece of (mis)information must first decide how to react—by openly endorsing it, remaining silent, or openly challenging it. After observing the reactions of her neighbors who also received the message, the agent then chooses whether to forward it to others in her own chatroom who have not yet received it. By distinguishing these two roles, our framework addresses puzzling real-world phenomena, such as the gap between what individuals privately believe and what they publicly transmit. A key assumption in our model is that, while perceived veracity influences decisions, the dominant factor is the alignment between an agent’s beliefs and those of her social network—a feature characteristic of communities formed through long-term offline relationships. This dynamic can lead agents to tacitly accept and even propagate information they privately judge to be of low credibility. Our results challenge the view that improving information literacy alone can curb the spread of misinformation. We show that when agents are highly sensitive to peer pressure and the network exhibits structural polarization, even if the majority does not genuinely believe in it, the message still can spread widely without encountering open resistance. 




\end{abstract}


\jelclassification{C72, D81, D83, D91}



\section{Introduction}\label{sec:int}

This paper investigates how information spreads within social networks composed of individuals who share long-term offline relationships. We focus on several empirical puzzles observed across social networks generally and examine how these dynamics operate in offline-based communities, with particular attention to their role in facilitating the diffusion of misinformation.


Growing political polarization and rapid advances in artificial intelligence have intensified concerns about misinformation, especially in the form of fake news. Although research on the drivers of misinformation diffusion has significantly expanded recently (e.g., \cite{vra18}; \cite{pr21}), most studies focus on open communication platforms (OCPs) such as Twitter, Facebook, and TikTok—partly because of their influence in the 2016 U.S. presidential election (e.g., \cite{ag17}; \cite{betal20}). By contrast, private messaging applications such as WhatsApp, WeChat, Telegram, and LINE remain relatively understudied, despite their crucial roles in global events ranging from the COVID-19 vaccination debates \citep{gpgl22} and the Russian invasion of Ukraine \citep{jr24} to network propaganda in Russia and Hungary \citep{oetal24}. These private networks often rival—and sometimes surpass—OCPs in spreading misinformation, making it essential to understand their underlying mechanisms.

We study how the diffusion of (mis)information in such private networks is endogenously shaped by agents’ strategic reasoning. 
The network is modeled as an ordered tree in which each agent sequentially assumes two distinct roles:\footnote{The root agent assumes only the second role, while each terminal agent assumes only the first.}
\begin{itemize}
\item \textbf{Receiver: Responding to a message}. Upon receiving a message from her immediate predecessor, the agent decides how to respond—by openly disapproving, remaining silent, or expressing approval. All immediate successors of the same predecessor respond simultaneously, and each observes the reactions of her peers.
\item \textbf{Sender: Transmitting a message}. After responding, the agent  decides whether to forward the message to her own immediate successors.
\end{itemize}

Empirical evidence suggests that responding and transmitting behaviors are often guided by different, even conflicting, criteria. Yet the two stages are interdependent: responses shape perceptions, which in turn affect transmission. Our model imposes a unifying decision principle across both roles: agents seek to maximize belief alignment within their chatrooms. But this principle operates differently in each role. As a receiver, an agent choose responses that signal a credence close to the chatroom’s average belief; as a (potential) sender, she transmits only if doing so is expected to bring her immediate successors’ beliefs closer to her own.

Our framework draws on psychological game theory (PGT, \cite{gps89, bd09, bd22, bcd19}) on social networks. PGT extends classical game theory by allowing utilities to depend on \emph{endogenous beliefs}.
As a receiver, an agent evaluates each possible response according to (i) the distance between the response’s implied credence and her own “real” credence (a measure of conscience), and (ii) the distance between the implied credence and others’ beliefs (peer pressure). A parameter captures her sensitivity to peer pressure. The smaller the combined distance, the higher the utility of that action.


After observing peers’ reactions, the agent assumes the role of a sender whose goal is to align others’ worldviews with her own. Here, by worldview we mean beliefs about fundamental hypotheses, such as political orientation or religious faith. Agents are assumed to share a consensus on evidence relationships—that is, a common understanding of how evidence bears on hypotheses.\footnote{For example, people across the political spectrum may disagree about whether a particular party is ethical or whether an allegation against its leader is true, but they can still agree that if the party were genuinely good, such an allegation would be unlikely.}  A sender benefits if transmitting the message may shift her successors’ posterior beliefs toward her worldview. 


In addition, the two stages are connected by a threshold: the agent only considers transmitting if, in the previous chatroom where she is a receiver, the number of explicit disapprovals remains below a given threshold. \medskip

This model unifies several insights that, to our best knowledge, have not previously been integrated into a single framework. First, in networks grounded in long-term offline interactions, peer pressure exerts a powerful influence. Even when individuals doubt a message’s credibility, they may remain silent—or even signal agreement—to preserve harmony. This feature plays an important role for the role of the receiver.

Second, existing research highlights three recurring features of senders’ behavior: (1)
Beliefs are often politically motivated \citep[e.g.,][]{ag17, pr19a, petal21};
(2) Individuals sometimes share information they neither scrutinize nor believe\citep[e.g.,][]{petal21, petal20};
(3) People nevertheless care about veracity \citep[e.g.,][]{pr19, petal21, dsa20}.

At first glance, these observations appear puzzling, if not contradictory. Our model, however, provides a consistent explanation. The first observation suggests a separation between an agent’s worldview and her credence in a specific claim. Yet people can have a consensus on evidence relationships. Viewed in this light, the second observation becomes easier to understand: even if a sender assigns low credence to a message, she may still transmit it because the message supports her worldview and may ``convince'' the receivers, that is, shift their posterior beliefs closer to her own.

The third observation can also be reconciled. When individuals lack direct means to verify information, they rely heavily on social cues as proxies for credibility. In particular, the absence of open refutation within their network may be interpreted as tacit endorsement, thereby lending the message perceived veracity and justifying its further transmission.\footnote{As noted by \cite{gs06}, access to reliable verification methods drastically alters individuals’ behavior.}


The solution concept we adopt is a modified Bayesian equilibrium. Each agent is endowed with a set of types, where each type represents her credence in the message. Also, each agent holds a  belief about others' types. Because they share a consensus on evidence relationships, these beliefs pushes forward a probability distribution over others’ views of the fundamental hypothesis. An equilibrium consists of a profile of action pairs—one for the role of receiver and one for the role of sender—such that:
\begin{itemize}
\item{\textbf{As receiver}.} For each of her types, an agent’s chosen response is a best reply to her belief about the credences of other chatroom members. That is, the action minimizes the combined distance of deviating from her own credence (conscience) and from others’ beliefs (peer pressure).

\item{\textbf{As sender}.} Given her belief about her immediate successors’ views on the fundamental hypothesis, an agent transmits the message if and only if (i) the number of disapprovals in the previous chatroom does not exceed the threshold, and (ii) sending the message reduces the distance between her successors’ belief about the fundamental hypothesis and her own.
\end{itemize}



The use of Bayesian equilibrium is standard in the literature on psychological game theory. \citet{bcd19} show that the solution concept introduced in the seminal paper by \citet{gps89} is in fact equivalent to Bayesian equilibrium. We therefore adopt Bayesian equilibrium in this tradition. Our formulation may not appear ``canonical'' because it requires that all types of an agent prescribe the \emph{same} best response. Yet this is only a matter of interpretation: an agent’s set of types reflects the uncertainty others hold about her, not uncertainty she holds about herself. From her own perspective, she has a fixed belief and thus makes a deterministic choice.

This definition also captures real-world dynamics in networks grounded in offline relationships. In such settings, the digital chatroom is less a new arena than an extension of established ties. Through repeated interactions, members develop a mental map of one another’s credence of a message as well as the ideological positions based on observed behavior, especially reactions to shared information. Over time, these expectations stabilize: participants’ beliefs about others converge to consistent patterns, which guides and stabilizes their actions. In this respect, our solution concept also carries the flavor of a self-confirming equilibrium \citep{bgm92, bg97, fl93, bpp23}.

We examine the conditions under which (mis)information reaches all agents in the network. Our first result shows that, holding other factors constant, a larger number of agents who openly disapprove of the message reduces the likelihood of further transmission. However, as agents’ sensitivity to peer pressure and the uncertainty increase, open disapproval becomes less likely.
Consequently, even individuals who privately doubt the message may remain silent or conform, enabling the spread of misinformation. This mechanism parallels findings on the persistence of harmful norms \citep[e.g.,][]{bgyd20}, and suggests that creating environments where individuals can express genuine views without fear of peer pressure may be more effective than focusing solely on information literacy—though such openness is especially difficult in hierarchical cultures.\footnote{Also, though websites and social media accounts / applications run by fact-checking and debunking institutions could provide a portable access to verify the factual accuracy of information, as pointed out in \cite{vra18}, it does not really effectively check the spreading of misinformation. Our model provides an explanation: in general, people do not care such an institution and assign a low sensitive parameter to it. Some private messaging application (for example, LINE) provides applications for spreading fact-checking information in chatrooms, whose effect might still be limited: it still needs an user to send it into the chatroom, and such behavior is equivalent to disapproval in our model, which is determined by one's beliefs about others and how much she care about them.}

Our second result focuses on the sender's behavior. Holding other factors constant, a potential sender is more likely to transmit a message when receivers who were initially skeptical become slightly more convinced, or when receivers who were already strong believers become even more engaged. By contrast, when receivers' beliefs are already close to the sender’s own, further increases in their credence reduce the sender's incentive to send. In short, greater like-mindedness discourages transmission.

If we assume that credences follow a bell-shaped distribution (e.g., normal), this result implies that polarization amplifies the spread of misinformation. Senders are most motivated to act when members are closer to the extremes, creating conditions that encourage further diffusion. This theoretical prediction is consistent with empirical findings on the relationship between polarization and the spread of misinformation online \citep[e.g.,][]{vetal16}.

Our findings highlight a limitation in current policies against misinformation. Many such policies assume that improving individual information literacy will curb misinformation. Our model challenges this optimism in the context of networks composed of long-term offline relationships. For instance, suppose the network tree is ordered by credence—so that in each chatroom the root assigns substantially higher credence to the message than most of her recipients. If each room contains a few true believers, while those with low credence are highly sensitive to peer pressure (as might be the case for younger members in hierarchical cultures), then misinformation will propagate through the entire network without resistance, even if the majority of agents do not genuinely believe it. As the title of the paper suggests, this outcome is even more troubling than Jonathan Swift’s famous observation: a lie can travel half way around while the truth never puts on its shoes.\footnote{This quotation is commonly attributed to Mark Twain. Yet according to recent research \citep{nc17}, the quotation first appeared in a line published by Jonathan Swift.}

\section{Preliminaries}

\subsection{A probability system of observable and unobservable events}\label{subsec:mod1}

Let $\Omega$ be an arbitrary set. Each $\omega \in \Omega$ represents a \emph{state of the world}, that is, a complete description of the world. For each $E \subseteq \Omega$, we use $\urcorner E$ to denote its complement, that is, $\urcorner E = \Omega \setminus E$. We fix a $\sigma$-algebra $\mathcal{F}$ on $\Omega$. Each $E \in \mathcal{F}$ is called an \emph{event}. We use $\Delta(\Omega)$ to denote the set of all probability measures on $\Omega$ with respect to $\mathcal{F}$. 

Among non-empty members of $\mathcal{F}$, we focus on two distinct events $C$ and $M$. The former, as well as its complementary $\urcorner C$, is a \emph{concealed} event, which is not directly observable, or at least not observable at the time a decision must be made. For example, $C$ could represents a value judgments such as ``Party A will improve social welfare and justice once it is elected''. The latter indicates a \emph{observable} or directly verifiable event. For example, $M$ could be ``Party A's leader had been arrested for drunk driving''. Each probability measure $\alpha \in \Delta(\Omega)$ represents a belief; especially, $\alpha(C)$ and $\alpha(M)$ represents the credence of $C$ and $M$, respectively. In our model, $\alpha(C)$ is interpreted as her belief about fundamental hypothesis, while $\alpha(M)$ is her assessment of the veracity of the specific message (evidence) $M$. We assume that the evidence relationship between $C$ and $M$ is fixed. Formally, we fix a pair $\overline{\mu} := \langle \overline{\mu}_{M|C}, \overline{\mu}_{M|\urcorner C}\rangle \in (0,1)\times (0,1)$; without loss of generality, we assume $\overline{\mu}_{M|C} > \overline{\mu}_{M|\urcorner C}$. We focus on $\alpha \in \Delta(\Omega)$ satisfying the following conditions:
\begin{itemize}
\item[\textbf{P1.}] $\alpha(C) \in (0,1)$ and $\alpha(M) \in (\overline{\mu}_{M|\urcorner C}, \overline{\mu}_{M|C})$;

\item[\textbf{P2.}] $\alpha (M|C) = \overline{\mu}_{M|C}$ and $\alpha(M|\urcorner C) = \overline{\mu}_{M|\urcorner C}$.
\end{itemize}
We use $\Delta^{\overline{\mu}, M^+}(\Omega)$ to denote the set of $\alpha \in \Delta(\Omega)$ satisfying P1 and P2. It can be seen that for each $\alpha \in \Delta^{\overline{\mu}, \mathcal{M}^+}(\Omega)$, the value of $\alpha(C)$ and $\alpha(M|C)$ are determined, namely,
\begin{equation}\label{eq:post}
\alpha(C)  = \frac{\alpha(M) - \overline{\mu}_{M|\urcorner C}}{ \overline{\mu}_{M|C} - \overline{\mu}_{M|\urcorner C}}, { \ \ }
 \alpha(C|M) = \frac{\overline{\mu}_{M|C}}{\alpha(M)}\times \alpha(C)
\end{equation}
Since $\alpha(M) \in (\overline{\mu}_{M|\urcorner C}, \overline{\mu}_{M|C})$, it is clear that $\alpha(C|M) > \alpha(C)$, which we say that \emph{$M$ favors $C$} (or, equivalently, \emph{$M$ does not favor $\urcorner C$}).

\subsection{The network of chatrooms as an ordered tree}
For simplicity, we assume that the social network of chatrooms form an ordered tree $\mathbf{T}=\langle\mathbf{I}, \mathbf{E} \rangle$, where $\mathbf{I}$ is the set of agents and $\mathbf{E}$ the set of directed edges. The tree starts from a root agent $\imath$, which could send the message $M$ to her immediate successors, and each of them could send $M$ to his or her own immediate successors, etc. The set of each non-terminal agent and her immediate successors form a \emph{chatroom}. 

\begin{figure}[htbp]
  \centering
  \tikzset{
    every node/.style={font=\large}, 
    edge from parent/.style={
      draw, -{Stealth[scale=1.0]}, semithick, shorten >=1pt
    },
    level 1/.style={sibling distance=50mm, level distance=18mm}, 
    level 2/.style={sibling distance=28mm, level distance=18mm}  
  }
  \begin{tikzpicture}[grow=down] 
    \node (1) {1}
      child { node (2) {2}
        child { node (5) {5} }
        child { node (6) {6} }
      }
      child { node (3) {3}
        child { node (7) {7} }
        child { node (8) {8} }
      }
      child { node (4) {4}
        child { node (9) {9} }
        child { node (10) {10} }
      };
  \end{tikzpicture}

  \caption{A directed rooted tree of chatrooms}
  \label{fig:dir-tree}
\end{figure}
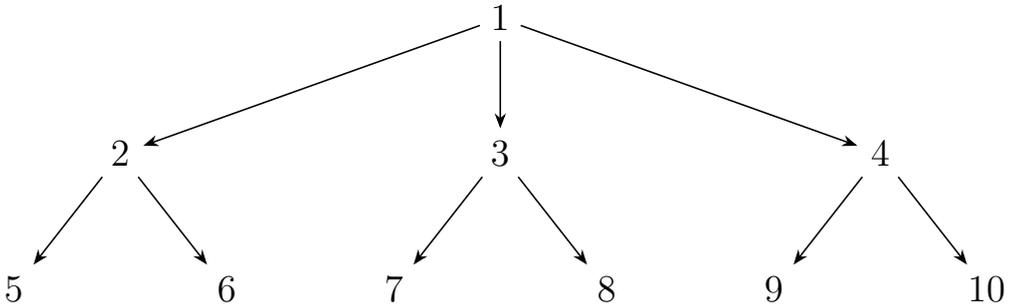

Each agent has to consider at least one chatrooms: the chatroom that she assumes the role of  receiver, and the chatroom that she assumes the role of sender. We use $\chat_R(i)$ to denote the set of all participants of former, and $\chat_S(i)$ the latter; when there is no need to specify the member, we use $\chat$ to denote an arbitrary chatroom. For each chatroom $\chat$, we use $\chat^\circ$ to denote the set of receivers in $\chat$. For example, in Figure \ref{fig:dir-tree}, $\chat_R(2) (=\chat_R(3) =\chat_R(4)= \chat_S(1))= \{1, 2, 3, 4\}$, $\chat_R^\circ(2) = \{2,3,4\}$.

An ordered tree is a useful simplification. First, they ensure that each chatroom has a unique source of information, eliminating conflicts from multiple origins. Second, they guarantee that every agent assumes one role in only one chatroom, making the analysis of strategic behavior more tractable.

\section{The role of receiver: Disapproval, silence, or approval}\label{m1}
\subsection{The model}

Suppose that a player $\iota \in I$ sends the message $M$ to her immediate successors. For a receiver $i \in \chat^\circ(\iota)$, she observes $\iota$'s dissemination of $M$ and she knows everyone in the chatroom received the $M$ simultaneously and she can observe their responses. Her payoff is henceforth influenced by two factors: her own and others' credence of $M$. For responding, she has three options: openly disapproving and denouncing that $M$ is misinformation, staying in silence, or openly approving $M$ to be true. We use $0$, $0.5$, and $1$ to denote the three actions, respectively; the rationale behind this nomenclature will be explained later. 

Agent $i$ has her credence of $M$, $\alpha_i(M) \in (\overline{\mu}_{M|\urcorner C}, \overline{\mu}_{M|C})$ (recall that $\alpha_i \in \Delta^{\overline{\mu}, M^+}(\Omega)$). She evaluate each action by the sum of its distance from her own credence and its weighted distance from the ``public opinion'' in the chatroom, and her objective is to minimize this distance. Formally, given a profile $(\alpha, a) = (\alpha_j(M), a_j)_{j \in \chat_R(i)}$, where for each $j \in \chat_R(i)$, $\alpha_j(M) \in (\overline{\mu}_{M|\urcorner C}, \overline{\mu}_{M|C})$ and $a_j \in \{0, 0.5, 1\}$, $i$'s (psychological) utility is 
\begin{equation}\label{puti}
u_i (\alpha,a) = -\left(|a_i -\alpha_i(M)|+\lambda_i\left|a_i -  \sum_{j\in \chat_R(i)\setminus \{i\}}\frac{\alpha_j(M)}{\# \chat_R(i)-1}\right|\right)
\end{equation}
Here, we use $\# E$ to denote the set $E$'s cardinality. The formula has two components,  representing two factors that represent the agent's decision The first, $|a_i - \alpha_i(M)|$, represents her conscience: it captures the discrepancy between her choice and her belief about $M$'s veracity. The larger this gap is, the more guilt she feels, as if she has betrayed her own genuine judgment.
The second, $\lambda_i\left|a_i -  \sum_{j\in \chat_R(i)\setminus \{i\}}\frac{\alpha_j(M)}{\# \chat_R(i)-1}\right|$, captures the influence of peer pressure on her. The average credence of her peers, $\sum_{j\in \chat_R(i)\setminus \{i\}}\frac{\alpha_j(M)}{\#\chat_R(i)-1}$, represents the public opinion in the chatroom; in the following, we will denote this value by $\mathsf{N}\left((\alpha_j(M))_{j \in\chat_R(i)\setminus \{i\}}\right)$. The \emph{sensitivity parameter} $\lambda_i \in [0,\infty)$ measures the extent to which the agent is influenced by peer pressure. Weighted by her sensitivity, the agent is also reluctant to deviate too far from public opinion, either to gain a sense of harmony or to avoid potential punishment. The agent chooses an action that balances the two factors, aiming to minimize their sum.

Some remarks are in order. One might wonder why the peer-pressure component depends on the agent’s action $a_i$ rather than her belief $\alpha_i(M)$. There are two reasons for this choice. First, it is true that what shapes an agent’s image in the eyes of her peers is not her action per se, but how her peers infer her belief from that action. The challenge, however, lies in specifying this inference. In principle, it could be endogenously determined, as in signaling models \citep{ck87}. To keep the model tractable, we instead assume that each response is interpreted at face value: disapproval implies that the agent treats $M$ as completely false (0), silence implies that she assigns equal probabilities to truth and falsity (0.5), and approval implies that she regards $M$ as completely true (1). 


Also, in many cases, the action itself carries meaning. As illustrated by classic examples—“calling a deer a horse” in the East and Caligula’s horse in the West—sometimes even when everyone’s genuine attitude is common knowledge, the action alone serves its purpose. In both examples, the courtiers knew that a deer was not a horse, or that a horse could not be a consul, and the rulers knew this as well. What mattered was not belief but performance: the courtiers’ echoing actions signaled their willingness to abandon reality in favor of obedience.

Second, one might ask whether we have to allow $A_i$ to take continuous values in $[0,1]$. While technically feasible, this extension would be neither realistic nor analytically illuminating for several reasons. First, real-world communication constraints make precise probability statements impractical. In everyday discourse, particularly in online chatrooms, nuanced probabilistic statements like ``I believe this news has a 28\% likelihood of being true'' are typically interpreted through a coarse categorical lens. Audiences tend to perceive such statements simply as agreement, disagreement, or uncertainty, effectively collapsing the continuous space into discrete categories.

Also, the continuous extension offers limited technical insights. The optimization problem's solution reduces to a simple binary choice to be either $\alpha_i(M)$ or $\mathsf{N}\left((\alpha_j(M))_{j \in\chat_R(i)\setminus \{i\}}\right)$ (or the interval with the two values as the ends when $\lambda_i=1$), determined solely by the value of $\lambda_i$. This eliminates much of the model's analytical richness.

\begin{example}\label{ex1}
Consider a situation where $i$ deems $M$ quite unlikely, say, $\alpha_i(M) = 0.1$. Suppose that  $\lambda_i =1$. Figure \ref{fig:UN01} (1) shows how the utility $u_i(a,\alpha)$ changes with respect to $\mathsf{N}\left((\alpha_j(M))_{j \in\chat_R(i)\setminus \{i\}}\right)$ for $a_i = 0, 0.5$, and $1$.

\begin{figure}[h!] 
\centering
  \includegraphics[width=1\columnwidth]{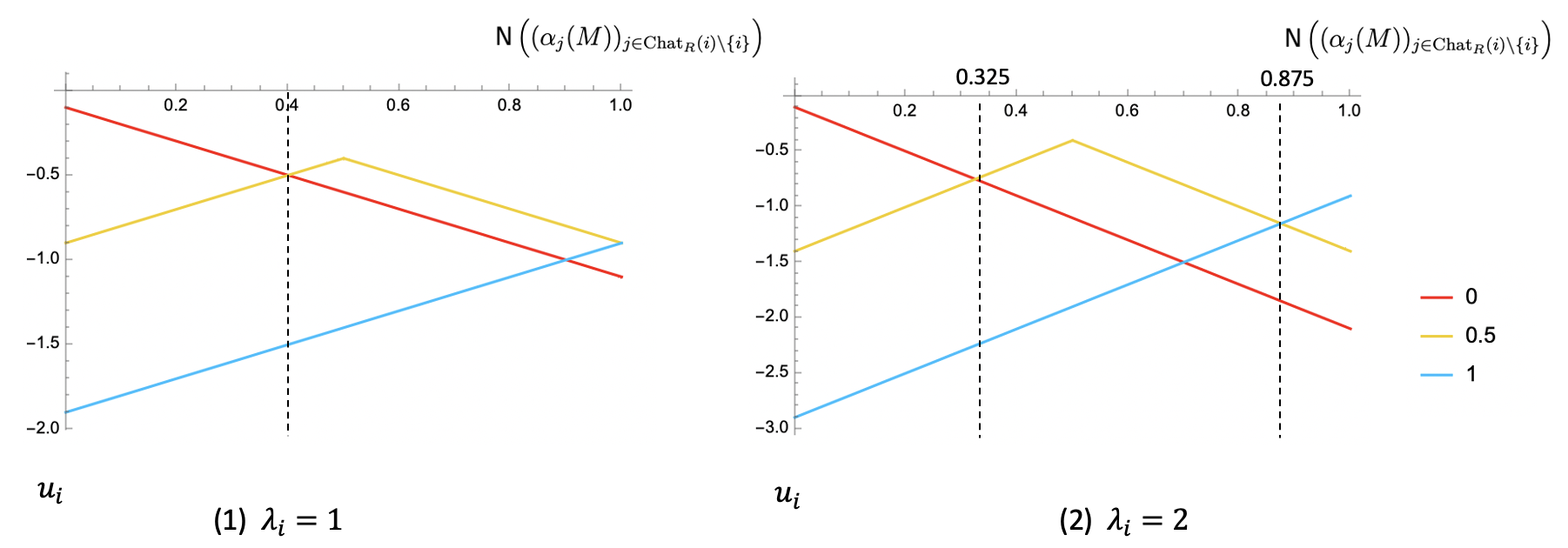}
  \caption{Best responses under two sensitivity parameters when $\alpha_i(M)= 0.1$}
  \label{fig:UN01}
\end{figure}

When $\mathsf{N}\left((\alpha_j(M))_{j \in\chat_R(i)\setminus \{i\}}\right) < 0.4$, openly disapproving is the optimal response; when $\mathsf{N}\left((\alpha_j(M))_{j \in\chat_R(i)\setminus \{i\}}\right) \in(0.4,1)$, remaining silent is preferable. When $\mathsf{N}\left((\alpha_j(M))_{j \in\chat_R(i)\setminus \{i\}}\right) =1$, both silence and approval are equally optimal. This result suggests that when sensitivity is not very high (recall that $\lambda_i = 1$ means she values the public opinion as much as hers), open approval is almost never optimal. In other words, agent $i$ rarely needs to bend so far as to betray her own judgment. Moreover, when she feels sufficiently assured—namely, when she believes that others also regard the information as having low veracity—she is able to express her opinion faithfully.


Yet when $\lambda_i =2$, since she now places greater importance on others' opinions of her, things will change. As shown in Figure \ref{fig:UN01} (2), for $\mathsf{N}\left((\alpha_j(M))_{j \in\chat_R(i)\setminus \{i\}}\right) > 0.875$, $i$ has to openly approve the information. It is straightforward to see that as $\lambda_i \rightarrow \infty$, the threshold that $i$ has to openly approve is $0.75$; also, one can see that the minimum upper bound for $i$ to refute $M$ is $0.25$. $\clubsuit$
\end{example}

\subsection{Some preliminary results}
Fix $M$. Let $b_i =\mathsf{N}\left((\alpha_j(M))_{j \in\chat_R(i)\setminus \{i\}}\right)$. For each $\lambda_i \in [0, \infty)$ and $a_i \in \{0, 0.5, 1\}$, let $\mathbb{A}_i(a_i; b_i, \lambda_i) : = \{\alpha_i(M):$ under $b_i, \lambda_i$ and $\alpha_i(M), a_i$ is an optimal choice$\}$. We have the following  result.

\begin{proposition}\label{re2}
For each $a_i$, $\mathbb{A}_i(a_i; b_i, \lambda_i)$ is either empty or a closed interval $[\underline{\alpha}_i(a_i;b_i, \lambda_i)$, $\overline{\alpha}_i(a_i; b_i, \lambda_i)]$. The following statements hold.

\begin{itemize}
\item[1] $\underline{\alpha}_i(0;b_i, \lambda_i) \leq \underline{\alpha}_i(0.5;b_i, \lambda_i) \leq$ $\overline{\alpha}_i(0; b_i, \lambda_i) \leq \underline{\alpha}_i(1;b_i, \lambda_i) \leq$ $\overline{\alpha}_i(0.5; b_i, \lambda_i) \leq \overline{\alpha}_i(1; b_i, \lambda_i)$. 

\item[2] For generic $\beta_i, \lambda_i)$, $\interior (\mathbb{A}_i(a_i; b_i, \lambda_i)) \cap \interior (\mathbb{A}_i(a_i^\prime; b_i, \lambda_i)) = \emptyset$ if $a_i \neq a_i^\prime$. Here, $\interior (E)$ denote the interior of $E$.

\item[3] If $\mathbb{A}_i(0.5; b_i, \lambda_i)) = \emptyset$, then either $\mathbb{A}_i(0; b_i, \lambda_i)) = \emptyset$ or $\mathbb{A}_i(1; b_i, \lambda_i)) = \emptyset$.

\end{itemize}

\end{proposition}

\begin{figure}[h!] 
\centering
  \includegraphics[width=1.1\columnwidth]{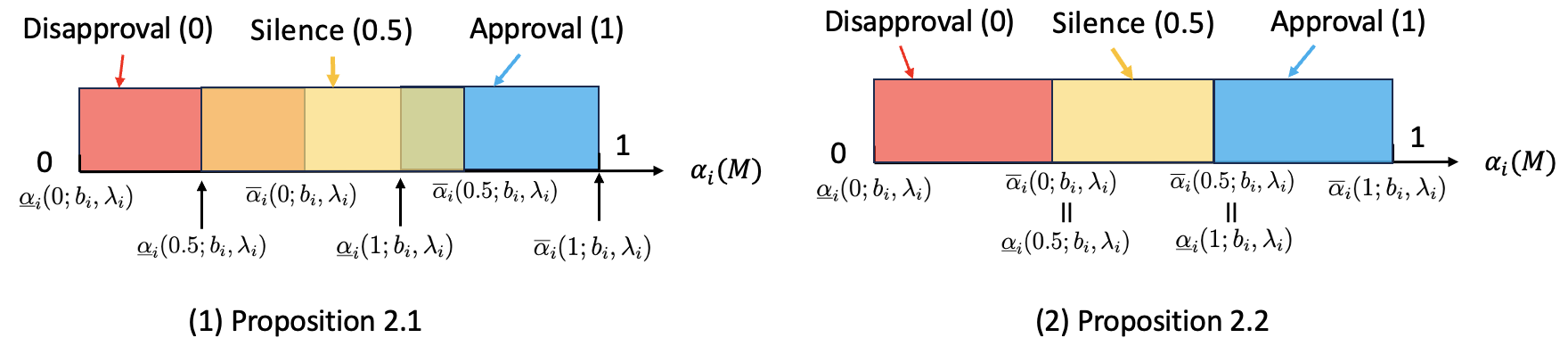}
  \caption{The general patterns in Proposition \ref{re2}, part one and part two}
  \label{fig:prop2}
\end{figure}

The idea underlying Proposition \ref{re2} is illustrated in Figure \ref{fig:prop2}. 

Proposition \ref{re2}.1 states that, given $\beta_i$ and $\lambda_i$, the set of $i$'s belief that support each action is a closed interval, and the locations of the three intervals are depicted in Figure \ref{fig:prop2} (1). In degenerate cases, two intervals may overlap and the intersection could even have a positive Borel measure. Yet in generic cases, the three intervals ``partition'' (if we ignore the overlapping at the end-points of intervals) the $[0,1]$ interval, as depicted in Figure \ref{fig:prop2} (2). Proposition \ref{re2}.3 could be taken as a result of ``immediate value'': if there are beliefs support approval and disapproval, there is some belief supporting silence.

Note that when we focus on $\mathsf{N}\left((\alpha_j(M))_{j \in\chat_R(i)\setminus \{i\}}\right)$, we implicitly assume that each agent's second-order belief on others' beliefs is Dirac (i.e., point-mass). Actually we can relax this assumption and generalize the result. Each agent can have a general belief $\beta^i$ which is a Borel probability measure on $(\overline{\mu}_{M|\urcorner C}, \overline{\mu}_{M|C})$. Since for each action $a_i$, $\mathbb{E}_{\beta^i} (|a_i - \mathsf{N}(\cdot)|)$ is a constant, one can see (from the proof) easily that the three-interval pattern still exists.



\begin{proof}[Proof of Proposition \ref{re2}]
1 and 2. Given $\lambda_i$ and $b_i$, one can see that the utility generated by three actions are as in Table \ref{TAB3}.
\begin{table}[ht!]
\centering
 \begin{tabular}{c c }
\hline
 Action   & $u_i(a, b_i)$ \\ 
 \hline
$0$ & $-\alpha_i(M) - \lambda_i b_i$  \\ 

$0.5$ & $-(\frac{1}{2} - \alpha_i(M) +\lambda_i |\frac{1}{2} - b_i|)$ if $\alpha_i(M) \leq \frac{1}{2}$, $-(\alpha_i(M) - \frac{1}{2} + \lambda_i |\frac{1}{2} - b_i|)$ if $\alpha_i(M) > \frac{1}{2}$ \\

$1$ & $\alpha_i(M) -1 - \lambda_i(1- b_i)$\\
 \hline
\end{tabular}
\caption{Utilities generated by each action}
\label{TAB3}
\end{table}

Given $b_i$ and $\lambda_i$, both $u_i(0, \alpha_i)$ and $u_i(1, \alpha_i)$ are both linear, with slope $-1$ and $1$, respectively; $u_i(0.5, \alpha_i)$ is a piece-wise linear function which has mirror symmetry at $\alpha_i(M) = \frac{1}{2}$, with slope $1$ at the left-hand side and $-1$ the right-hand side. Hence, the optimal choice is determined by the relationship between the intercepts of the three functions. 

To see it, without loss of generality, suppose that $b_i \leq \frac{1}{2}$. One can see that the intercept of $u_i(0, \alpha_i)$ is $-\lambda_ib_i$ (for simplicity, we denote this value by $A$), that of $u_i(0.5, \alpha_i)$ is $-\frac{1}{2} - \lambda_i(\frac{1}{2} - b_i)$ (denoted by $B$), and that of $u_i(1, \alpha_i)$ is $-1 - \lambda_i(1- b_i)$ (denoted by $C$). It is easy to see that $B > C$. Hence, the relationship of the three functions belongs to one pattern in Figure \ref{fig:prop2proof}. One can easily see that the statement holds in each case. In a similar manner we can show that the statement holds when $b_i \geq \frac{1}{2}$.

\begin{figure}[h!] 
\centering
  \includegraphics[width=1\columnwidth]{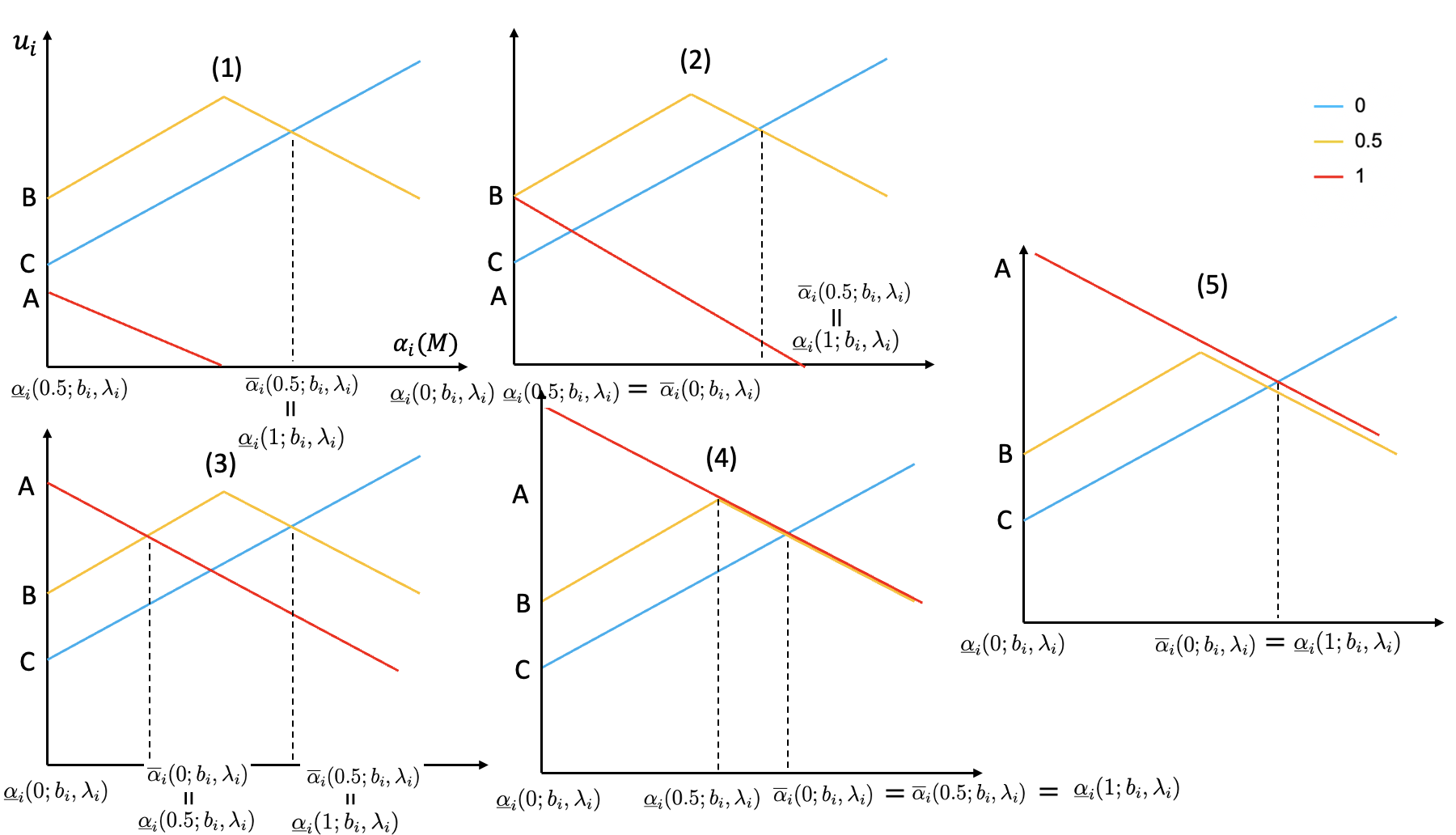}
  \caption{Three patterns of the relationship among the utility functions of each actions}
  \label{fig:prop2proof}
\end{figure}

3. Without loss of generality, we still assume that $b_i \leq \frac{1}{2}$. To show the statement holds, we need to show that if the pattern depicted in Figure \ref{fig:prop2proof} (5) happens, that is, $u_i(0, \alpha_i(M))$ does not cross (i.e., $\mathbb{A}(0.5; b_i, \lambda_i) = \emptyset$), then $u_i(0.5, \alpha_i(M))$ and $u_i(0, \alpha_i(M))$ crosses $u_i(1, \alpha_i(M))$ for some $\alpha_i(M) \in [0,1]$. Formally, we have to show that the following two linear inequalities are inconsistent:
\begin{equation*}
u_i\left(0, \frac{1}{2}\right) > u_i\left(0.5, \frac{1}{2}\right), \text{ and } u_i(0,1) \leq u_i(1,1)
\end{equation*}
The first inequality implies that $\lambda_i \neq 0$ and $b_i < \frac{1}{4}(1-\frac{1}{\lambda_i})$, while the second one implies either $\lambda_i = 0$ or $b_i \geq \frac{1}{2}(1- \frac{1}{\lambda_i})$. Since both $b_i$ and $\lambda_i$ have to be non-negative, the two implications are inconsistent. Therefore, the third case in Figure \ref{fig:prop2proof} can never happen. In a similar manner we can show the statement holds when $b_i > \frac{1}{2}$.
\end{proof}

\begin{proposition}\label{prop:extension}
Suppose that $a_i^*$ is the closest action to $b_i$. 
\begin{enumerate}
\item Then for each $\lambda_i, \lambda_i^\prime$ with $\lambda_i \leq \lambda_i^\prime$, $\mathbb{A}_i(a_i^*;b_i, \lambda_i) \subseteq \mathbb{A}_i(a_i^*;b_i, \lambda_i^\prime)$.

\item Of $a_i^*$ is the unique closest action to $b_i$, then for each $\alpha_i(M)$, there is some $\lambda_i$ such that $\alpha_i(M) \in \mathbb{A}_i(a_i^*;b_i, \lambda_i)$.
\end{enumerate}
\end{proposition}

It is easy to generalize the result: if there are two actions that are closest (or closer than the third action) to $b_i$, then for each $\alpha_i(M)$ there is some $\lambda_i$ such that $\alpha_i(M) \in \mathbb{A}_i(a_i^*;b_i, \lambda_i)$, where $a_i^*$ is the action among the two closest actions to $b_i$ which is closer to $\alpha_i(M)$. To summarize, the statement implies that, as one's sensitivity to the peer pressure increases, her own credence of $M$ will be irrelevant to her choice.

\begin{proof}[Proof of Proposition \ref{prop:extension}]
1. Let $a_i \neq a_i^*$. We need to show that if $|a_i^*-b_i| \geq |a_i-b_i|$ and
\begin{equation}\label{lambda1}
|a_i^*-\alpha_i(M)| + \lambda_i|a_i^*-b_i| \leq |a_i-\alpha_i(M)| + \lambda_i|a_i-b_i|,
\end{equation}
then for each $\lambda^\prime_i > \lambda_i$,
\begin{equation}\label{lambda2}
|a_i^*-\alpha_i(M)| + \lambda^\prime_i|a_i^*-b_i| \leq |a_i-\alpha_i(M)| + \lambda^\prime_i|a_i-b_i|
\end{equation}
Suppose that for some $\lambda^\prime_i > \lambda_i$, inequality (\ref{lambda2}) does not hold, i.e., 
\begin{equation}\label{lambda3}
|a_i^*-\alpha_i(M)| + \lambda^\prime_i|a_i^*-b_i| > |a_i-\alpha_i(M)| + \lambda^\prime_i|a_i-b_i|
\end{equation}
To simplify symbols, we use $A$ to denote the value of $|a_i^*-\alpha_i(M)|$, $B$ for $|a_i^*-b_i|$, $C$ for $|a_i-\alpha_i(M)|$, and $D$ for $|a_i-b_i|$. Then, by our assumption, $B-D \leq 0$. It follows from (\ref{lambda1}) that
\begin{equation}\label{lambda11}
\lambda_i(B-D) \leq C-A
\end{equation}
and 
\begin{equation}\label{lambda31}
\lambda^\prime_i(B-D) > C-A
\end{equation}
Here we need to discuss several cases. First, if $\lambda_i = 0$, then $C-A \geq 0$. Yet by (\ref{lambda31}) because, since $\lambda_i^\prime >0$, it follows that $B-D > 0$, which is incompatible to our assumption that $B - D \geq 0$. 

Second, suppose that $\lambda_i > 0$. In this case, if $C-A \leq 0$, we can directly see the contradiction. Indeed, by (\ref{lambda11}), it follows that $B-D \leq \frac{C-A}{\lambda_i}$; yet by (\ref{lambda31}), it follows that $B-D > \frac{C-A}{\lambda^\prime_i} \geq  \leq \frac{C-A}{\lambda_i} \geq B-D$. On the other hand, if if $C-A > 0$, we return to the incompatibility in the first case: $\lambda^\prime_i(B-D) \geq 0$, and (\ref{lambda31}) cannot hold.

2. From the argument above, it is easy to see that, given $\alpha_i(M)$, if the condition
\begin{equation*}
\lambda_i \geq \max_{a_i \in A_i\setminus \{a_i*\}}\left\{0, \frac{|a_i - \alpha_i(M)|-A}{B-|a_i - b_i|} \right\}
\end{equation*}
 is satisfied, $\alpha_i(M) \in \mathbb{A}_i(a_i^*;b_i, \lambda_i)$.
\end{proof}

\subsection{Solution concept in a chatroom}
We adopt Bayesian equilibrium as the solution concept. For each chatroom $\chat$, the reaction-to-message-$M$ situation can be described as a Bayesian game $\Gamma(\chat, $ $\lambda, \Theta, p) = \langle I, (\Theta_i, A_i, u_i, p^i)_{i \in I}\rangle$ where
\begin{itemize}
\item $I = \chat$; 

\item For each $i \in I$, 
\begin{itemize}

\item $\Theta_i \subseteq (\overline{\mu}_{M|\urcorner C}, \overline{\mu}_{M|C})$,

\item $A_i = \{$WAIT$\}$ if $i$ is the root of $\chat$, and $A_i = \{0, 0.5, 1\}$ if $i \in \chat^\circ$,

\item For each $\theta = (\theta_j)_{j \in I} \in \Theta$ ($:= \times_{j \in I}\Theta_j$) and $a = (a_i)_{i \in I} \in A$ (: = $ \times_{j \in I}A_j$), 
\begin{equation}\label{repe}
u_i (a, \theta) = -|a_i - \theta_i|-\lambda_i\left|a_i -  \sum_{j\in I\setminus \{i\}}\frac{\theta_j}{\#I-1}\right|
\end{equation}

\item $p^i \in \Delta(\Theta_{-i})$.
\end{itemize}
\end{itemize}

Here, each type $\theta_i \in \Theta_i$ represents $i$'s credence of $M$, i.e., $\theta_i = \alpha_i(M)$ in the previous subsections, which is her first-order belief. The probability $p^i$ represents an distribution of other members' beliefs about veracity of $M$, which describes $i$'s second-order belief. 
Note that $\Theta_i$ could be a proper subset of $ (\overline{\mu}_{M|\urcorner C}, \overline{\mu}_{M|C})$; this is not trivial since, as a part of the rule of the game, it is assumed to be commonly known. One will see its critical role in condition E1 in the following Definition \ref{d1} and  in Example \ref{exxx2}.


One can also see that the root agent is not an active player in the game: she has only one available action—to WAIT for others’ responses. Consequently, her belief is irrelevant. However, her type still plays a role in the receivers’ beliefs, as it forms part of the public opinion.

\begin{definition}\label{d1}
Consider the Bayesian game $\Gamma(\chat, \lambda, \Theta, p)$. A \emph{(chatroom) equilibrium} is a profile of choice functions $(s_i: \Theta_i \rightarrow A_i)_{i\in \chat}$ satisfying

\begin{itemize}

\item[\textbf{E1}.] For each $i$ and each $\theta_i, \theta_i^\prime \in \Theta_i$, $s_i(\theta_i) = s_i(\theta_i^\prime)$;

\item[\textbf{E2}.] For each $i \in \chat^o$ and each $\theta_i \in \Theta_i$, $s_i(\theta_i) \in \arg \max_{a_i \in A_i} \mathbb{E}_{p^i}(u_{i} (a_i, a_{-i}), \theta_i, \cdot))$, where $a_{-i} = (a_j)_{j \neq i} = (s_j(\theta_j))_{j \neq i}$ for each $(\theta_j)_{j \neq i} \in \times_{j \neq i} \Theta_j$.
\end{itemize}

\end{definition}




\begin{example}\label{exxx2}
Consider a chatroom $\chat$ with two receivers, $1$ and $2$, to whom the root agent $\imath$ sends the message $M$. The sensitivity parameter $\lambda_1 = \lambda_2 = 3$. Consider a simple case where for each $i \in \{1,2\}$, her belief is a Dirac measure which assigns probability one to the point-event that each of the other two agents in the chatroom gives credence $0.8$ to $M$. Is there any chatroom equilibrium in which both $1$ and $2$ choose 0.5? The answer is no. Given that for each $i$, $\lambda_i = 3$, $\mathsf{N}\left((\theta_j)_{j \in\chat_R(i)\setminus \{i\}}\right)= 0.8$ and $s_i(\theta_i)= 0.5$ for each $\theta_i \in \Theta_i$, it follows that $\Theta_i \subseteq [0,0.6]$, for if $\theta_i \in (0.6,1]$, the choice can only be $1$. Yet if so, it is impossible to have the measure on $\Theta_{-i}$ which assigns one to $(0.8, 0.8)$.  Actually, one can find that in this case, ``everyone choosing $0.5$ in the equilibrium'' and ``$p^i((0.8, 0.8))= 1$ for $i = 1,2$'' are incompatible; we either have to sacrifice the former, for example, by letting $\Theta_i \subseteq [0.6,1]$ and let the equilibrium satisfy $s_i(\theta_i) = 1$ for each $i = 1,2$, or the latter, for example, by supposing the support of each $p^i$ to be close to $0.5$.

To illustrate an equilibrium, keep $p$ and suppose $\lambda_i = 6$ for each $i$. In this case, for each $\theta_i \in [0,1]$,  the optimal action is $1$. One can easily see that in this case, any $\Theta$ with $0.8 \in \Theta_i$ for each $i$ support an equilibrium, in which both $1$ and $2$ choose $1$, regardless of their genuine beliefs. Note that in this case, even though every one may deem the information rather unlikely, since they believe that others would assign it a relatively high likelihood, they choose to approve it; worse, they cannot revise their belief by observing others' choices. $\clubsuit$


\end{example}

In Example \ref{exxx2}, one might notice an interesting phenomena: As each agent's sensitivity increases, her own belief matters less, and the public opinion becomes harder to resist and finally will dominates the choice. This intuition is captured by the following proposition.



\begin{proposition}
Fix $\chat$ and $p = (p^i)_{i \in \chat}$ with each $p^i$ a Dirac measure such for each $i, j \in \chat^\circ$, the average credence of her peers are the same, which is denoted by $b_i$.
Let $a$ be the option among $0, 0.5, 1$ which is closest to $b_i$. There is $\lambda^*$ such that
\begin{enumerate}
\item If $\lambda_i \geq \lambda^*$ for each $i\in I$, $\Gamma(\chat, \lambda, (\overline{\mu}_{M|\urcorner C}, \overline{\mu}_{M|\urcorner C})^{\chat}, p)$ has an equilibrium;

\item If $\lambda_i > \lambda^*$ for each $i\in I$, for each game $\Gamma(\chat, \lambda, \Theta, p)$ whose equilibrium set is non-empty and for each of its equilibrium $s$, $s_i(\theta_i) = a$ for each $i$ and each $\theta_i \in \Theta_i$.

\end{enumerate}

\end{proposition}
The result can be easily generalized: $p^i$ does not necessarily be a Dirac, and the average credence of agents do not have to be the same. The point is, for each $i$, if action $a^*_i$ minimize $\mathbb{E}_{p^i}(|a_i - \mathsf{N}(\cdot)|)$, this action will become dominant as $\lambda_i$ grows larger, and finally the minimizer will be (generically) the unique choice in each equilibrium.

This result can be interpreted as follows: as one cares more about the peer pressure (which seems to be more serious in a chatroom formed by acquaintances that those by strangers), finally, one's choice has nothing to do with her own belief, and one's belief cannot be refuted by observation.

\begin{proof}
 Without loss of generality, suppose that among all actions in $A_i$, $0$ is the closest to $b_i$, or, equivalently, $b_i < \frac{1}{4}$. Given $\theta_i \in [0,1]$,  the utility generated by three actions are as in Table \ref{TABassm1}.
\begin{table}[ht!]
\centering
 \begin{tabular}{c c }
\hline
 Action   & $u_i(a, b_i)$ \\ 
 \hline
$0$ & $-\theta_i - \lambda_i b_i$  \\ 

$0.5$ & $-(\frac{1}{2} - \theta_i +\lambda_i |\frac{1}{2} - b_i|)$ if $\theta_i \leq \frac{1}{2}$, $-(\theta_i - \frac{1}{2} + \lambda_i |\frac{1}{2} - b_i|)$ if $\theta_i > \frac{1}{2}$ \\

$1$ & $\theta_i -1 - \lambda_i(1- b_i)$\\
 \hline
\end{tabular}
\caption{Utilities generated by each action}
\label{TABassm1}
\end{table}

We have to discuss three cases. First, when $\theta_i \in [0, \frac{1}{2}]$, one can see that $0$ is the best option if $\lambda_i \geq \max\left\{0, \frac{2\theta_i -1}{1-4b_i}\right\} $. Second, when $\theta_i \in (\frac{1}{2}, 1]$, one can see that  $0$ is the best option if $\lambda_i \geq \frac{1}{1-4b_i}$. Therefore, by letting $\lambda^* = \frac{1}{1-4b_i}$, one can see that for each $\lambda$ with $\lambda_i > \lambda^*$, in each equilibrium $s_i(\theta_i) = 0$, and there is an equilibrium where $\Theta_i=[0,1]$.
\end{proof}




\section{The role of sender: Disseminate or not}\label{stage2}


\subsection{The model}

Recall that $\chat_S(i)$ is the chatroom where $i$ assumes the role of sender. Given the profile of first-order beliefs $\alpha=(\alpha_j)_{j\in \chat_S(i)}$, player $i$'s (psychological) utility for belief-similarity is defined as
\begin{equation}\label{psu}
\tilde{v}_i(\alpha) = \sum_{j \in \chat^\circ_S(i)} |\alpha_j(C) - \alpha_i(C)|
\end{equation}
This is the status quo of ``the similarity of worldview'' or ``like-mindedness''.Actually, this distance is equivalent to the taxicab distance defined on $\{C, \urcorner C\}$, which can be interpreted as measuring the similarity between agents’ worldviews under a binary fundamental hypothesis. By transmitting $M$, agent $i$ may change the the world view of her immediate successors, and the updated similarity is
\begin{equation*}
\sum_{j \in \chat_S^\circ(i)} |\alpha_j(C|M) - \alpha_i(C)|
\end{equation*}

If not transmitting, $i$'s payoff is $0$. Therefore, $i$'s decision depends on
\begin{equation*}\label{change}
v_i(\alpha):=\tilde{v}_i(\alpha) -  \sum_{j \in \chat^\circ_S(i)} |\alpha_j(C|M) - \alpha_i(C)|
\end{equation*}
is positive or not. 
That is, if $M$ decreases the gap between others' posterior belief and $i$'s \emph{original} belief (i.e., making others' faith closer to her own), transmitting $M$ brings positive psychological value to $i$. This corresponds to the observation that people are politically motivated. On the other hand, note that $i$ does not update her own belief according to $M$, that is, $i$ herself could be immune to the piece of information. This corresponds to the observation that people usually do not pay enough attention to the message they transmit.\footnote{When studying other issues like echo chamber and radicalization, updating one's own belief could be important.} 



To connect withe the previous chatroom where the sender is a receiver, we consider the network structure. When an agent has no effective way to confirm the veracity, the only evidence that she can refer to is the ``collective wisdom'', that is, the reaction of her fellow correspondents of the message. Therefore, an intuitive way is to assume that agent $i$ has a threshold $\ell_i$; as long as the number of agents in her network who openly disapproved the information is less than $\ell_i$, maintains a relatively high level of confidence in it.\footnote{One need to pay attention to the way of counting here. For example, when $i$ transmits the message only if  $0$ agent refuted it, $\ell_i $ is $1$, not $0$.} Here, confidence indicates agent $i$’s belief about the degree to which recipients will “buy” the message—that is, how likely, and to what extent, they will accept it.\footnote{For example, she may feel embarrassed if a significant number of people question the information after she has transmitted it. In this sense, she can be regarded as using her recipients’ reactions within the network as a sample for her statistical inference.} 


Recall that $\chat_R(i)$ is the chatroom where $i$ received the message $M$, and $a_{\chat^\circ_R(i)}$ be the vector of her and her fellow recipients' reactions, we use $\#_{a_{\chat^\circ_R(i)}}(0)$ to denote the number of agents in $\chat^\circ(\imath)$ who chose to openly refute it (i.e, chose action $0$), i.e., $\#_{a_{\chat^\circ_R(i)}}(0) : = \# \{j \in \chat^\circ_R(i): a_j = 0\}$. The threshold could be described as
\begin{equation}\label{tre}
\max \{\ell_i - \#_{a_{\chat^\circ_R(i)}}(0), 0\}
\end{equation}

Combining (\ref{change}) and (\ref{tre}), given a profile of prior $\alpha = (\alpha_j)_{j \in \chat_S(i)}$, we obtain an agent's (psychological) payoff function for sending a piece of information $M$:
\begin{equation}\label{send}
\tag{\textbf{SEND}}
\begin{split}
V_i(\alpha) &= \max \{\ell_i - \#_{a_{\chat^\circ_R(i)}}(0), 0\} \times v_i(\alpha)\\
& = \max \{\ell_i -\#_{a_{\chat^\circ_R(i)}}(0), 0\} \times \left(\tilde{v}_i(\alpha_i) -  \sum_{j \in \chat_S^\circ(i)}|\alpha_j(C|M) - \alpha_i(C)|\right)
\end{split}
\end{equation}

Recall that if $i$ does not send $M$, her payoff is 0. To break the tie, we assume that when by sending the message $V_i(\alpha) = 0$, $i$ will not send the message.


As an illustration, consider the following example. 
\begin{example}\label{ex3}
Suppose that $\overline{\mu} = \langle 0.01, 0.02 \rangle$. That is, for each agent $i$,
\begin{equation*}
\alpha_i(M|C) = 0.02, \ \alpha_i(M|\urcorner C) = 0.01
\end{equation*}

Consider the network in Figure \ref{fig:two-trees} (1). Suppose that $\alpha_1(C) = 0.7$, $\alpha_2(C) = 0.9$, and $\alpha_3 (C) = 0.2$; in other words, in 1's chatroom, agent 2's has similar \emph{world view} to 1's (i.e., both assign $C$ with a higher likelihood) while 3's is different from them. Now suppose the number of disapproval does not pass the threshold of $1$ in the previous stage. Should agent $1$ transmit it?

\begin{figure}[htbp]
  \centering

  \begin{minipage}{0.45\textwidth}
    \centering
    \begin{tikzpicture}[
      every node/.style={font=\large},
      edge from parent/.style={draw, -{Stealth[scale=1.0]}, semithick, shorten >=1pt},
      level 1/.style={sibling distance=35mm, level distance=15mm}
    ]
      \node {1}
        child {node {2}}
        child {node {3}};
    \end{tikzpicture}

    (1)
  \end{minipage}
  \hfill
  \begin{minipage}{0.45\textwidth}
    \centering
    \begin{tikzpicture}[
      every node/.style={font=\large},
      edge from parent/.style={draw, -{Stealth[scale=1.0]}, semithick, shorten >=1pt},
      level 1/.style={sibling distance=15mm, level distance=15mm}
    ]
      \node {1}
        child {node {2}}
        child {node {$2^\prime$}}
        child {node {$2^{\prime \prime}$}}
        child {node {3}};
    \end{tikzpicture}

    (2)
  \end{minipage}

  \caption{Two chatrooms.}
  \label{fig:two-trees}
\end{figure}
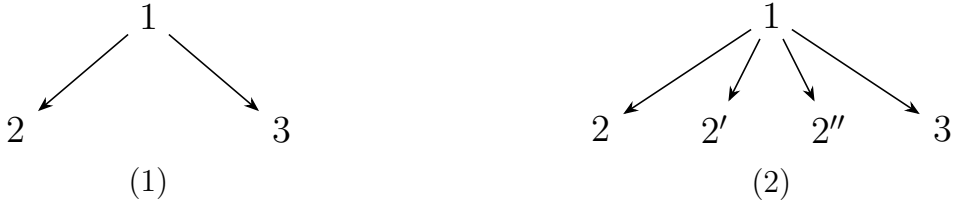

First, one can see that 
\begin{equation*}
\tilde{v}_1(\alpha) = |0.9-0.7|+|0.2-0.7| = 0.7
\end{equation*}
Agent 1 transmits the message only if it makes the posterior beliefs of agents in $\chat(1)$ on $\mathcal{C}$ closer to hers. One can see that $\alpha_2(C|M) = \frac{18}{19}$ and $\alpha_3(C|M) = \frac{1}{3}$, and consequently,
\begin{equation*}
\begin{split}
 \sum_{j \in \chat^\circ(i)} \sum_{C \in \mathcal{C}}|\alpha_j(C|M) - \alpha_i(C)| &=  \left|\frac{18}{19}-\frac{7}{10}\right| + \left|\frac{1}{3} - \frac{7}{10}\right| \\ &= \frac{35}{57} \thickapprox 0.614 <0.7
 \end{split}
\end{equation*}
Hence $v_1(\alpha) >0$ and agent $1$ should transmit $M$ to her chatroom.

Note that here, transmitting $M$ has is a double-edged sword. On one hand, it brings agent 3's belief closer to 1's, which is admirable; on the other hand, it will also ``radicalize'' agent 2 and makes the posterior of 2 farther from 1's, which is not admirable. Yet here, the former dominates the latter, which makes transmitting the better choice. One can see that if there are three replicas of $2$ as in Figure \ref{fig:two-trees} (2), $V_1(\alpha) <0$, and 1 should not transmit the information. $\clubsuit$
\end{example}


\section{Combining the two stages: integrating two belief systems}\label{sec:stable}
\subsection{The solution concept}
In this section, we provide a general framework to combine the two aforementioned stages. Based on the equilibrium defined in Definition \ref{d1}, a stable state can be defined, which can be used for static analysis. 


Fix an ordered tree $\mathbf{T} = \langle\mathbf{I}, \mathbf{E}\rangle$ with $\imath \in \mathbf{I}$ as the root, and a profile $\langle (\Theta_i, \lambda_i, p^i)_{i \in \mathbf{I}} \rangle$ where for each $i \in \mathbf{I}$, $\Theta_i$ is a Borel subset of  $(\overline{\mu}_{M|\urcorner C}, \overline{\mu}_{M|C})$, $\lambda_i \in [0, \infty)$, and 
\begin{itemize}
\item $p^\imath = p^\imath_S \in \Delta(\times_{j \in \chat_S^\circ(\imath)}\Theta_j)$,

\item For each non-terminal $i$ with $i \neq \imath$, $p^i = \langle p^i_R, p^i_S \rangle$ such that 
\begin{itemize}
\item $p^i_R \in \Delta(\times_{j \in \chat_R(i)\setminus \{i\}}\Theta_j)$,

\item $p^i_S \in \Delta(\times_{j \in \chat_S^\circ(i)}\Theta_j)$.
\end{itemize}
\item For terminal $i$, $p^i = p^i_R \in \Delta(\times_{j \in \chat_R(i)\setminus \{i\}}\Theta_j)$.

\end{itemize}

Consider a vector $(a_i)_{i \in \mathbf{I}}$ of action such that
\begin{itemize}
\item $a_\imath = a_\imath^S \in \{$S, NS$\}$;

\item For each non-terminal-non-root $i$, $a_i = (a_i^R, a_i^S) \in \{0,0.5,1\} \times  \{$S, NS$\}$;

\item For each terminal root $i$, $a_i = a_i^R \in \{0,0.5,1\}$.
\end{itemize}

We say that $(a_i)_{i \in \mathbf{I}}$ is an \emph{(global) equilibrium} if the following conditions are satisfied: In each chatroom $\chat$, 
\begin{itemize}
\item For the root $\imath$, $a_\imath^S =$ S if and only if for each $\theta_\imath \in \Theta_\imath$, if and only if for each $\theta_\imath \in \Theta_\imath$, $\mathbf{E}_{p_S^\imath}\left(v_\imath(\theta_\imath, \cdot\right)) >0$;

\item For the non-terminal $i$ with $i \neq \imath$, $a_i^S =$ S if and only if for each $\theta_i \in \Theta_i$, $\mathbf{E}_{p_S^i}\left(V_i(\theta_i, \cdot\right)) >0$;

\item If $a_{i} =$S, the profile $(s_j)_{j \in \chat}$, where $s_j(\theta_j) = a^R_j$ for each $j \in \chat^\circ$ and each $\theta_j \in \Theta_j$, is a local equilibrium of the game $\Gamma(\chat, \lambda_{\chat}, \Theta^{\chat}, (p^j_R)_{j \in \chat^\circ})$.

\end{itemize}

We use $GE(\mathbf{T}, (\Theta_i, \lambda_i, p^i)_{i \in \mathbf{I}})$ to denote the set of global equilibrium. 


\subsection{An example}\label{sec:eq-ex}
\begin{figure}[h!] 
\centering
  \includegraphics[width=1\columnwidth]{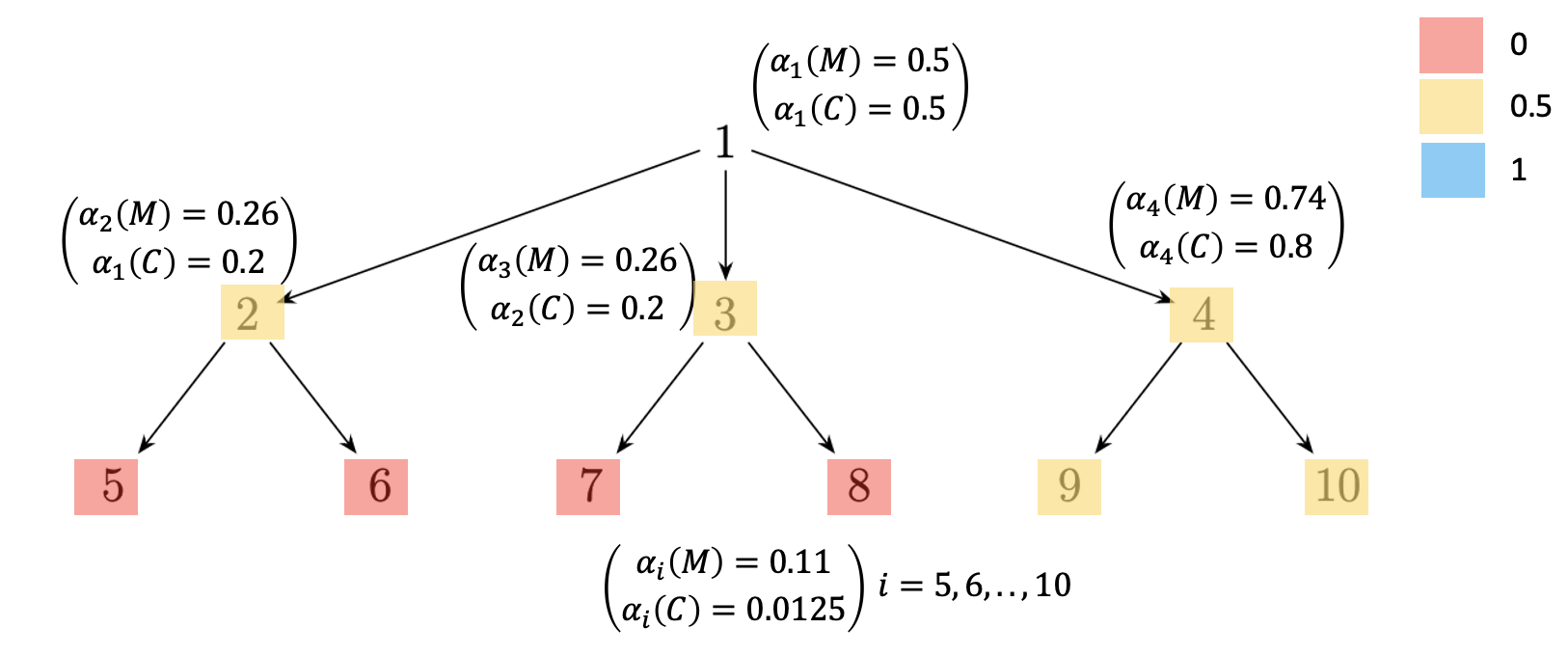}
  \caption{A network of chatrooms without uncertainty}
  \label{fig:FULL}
\end{figure}

Consider the tree in Figure \ref{fig:dir-tree}. We start from the simplest case where there is no uncertainty about each agent's belief about the veracity of $M$ (that is, for each $i$, $\Theta_i$ is a singleton), $\lambda_i = 1$, and $\ell_i=1$ or each $i$ (i.e., an agent will transmit the message, in case that she believes it would make the receivers' world view closer to hers, if no agent openly disapproves it). In addition, we assume that $\overline{\mu}_{M|C} = 0.9$ and $\overline{\mu}_{M|\urcorner C} = 0.1$.The numerical value of each agent's type $\theta_i = \alpha_i(M)$ is listed in Figure \ref{fig:FULL}; based on $\overline{\mu}$, we can compute (the unique value) $\alpha_i(C)$ for each $i$. 

For example, for agent $1$ (the root), $\Theta_1 = \{\theta_1\} = \{0.5\}$; it follows that $\alpha_1(C) = 0.5$. For agents $2$ and $3$, since $\theta_2 = \theta_3$ ($= \alpha_2(M) = \alpha_3(M) = 0.26$), one can see that $\alpha_2(C) = \alpha_3(C) = 0.26$.

There is a unique equilibrium, in which every non-terminal agent chooses to send the message. Indeed, for example, agent 1 choose to send $M$ because 
\begin{equation*}
\begin{split}
\tilde{v}_1(\alpha_1) &= |0.5 - 0.2|\times 2 \times 2 + |0.8-0.5|\times 2 = 1.8\\
\alpha_2(C|M) & = \alpha_3(C|M)  \thickapprox 0.5625, { \ \ } \alpha_4(C|M) \thickapprox 0.9844,\\
\sum_{j \in \chat^\circ(1)} |\alpha_j(C|M) - \alpha_1(C)| & \thickapprox |0.5625 - 0.5|  \times 2 + |0.9844 - 0.5|  = 0.6094
\end{split}
\end{equation*}
It follows that
\begin{equation*}
\begin{split}
V_1(\alpha) &= \max \{\ell_1 - \#_{a_{\chat^\circ}}(0), 0\} \times v_1(\alpha)\\
 &   = \max \{\ell_1 -\#_{a_{\chat^\circ}}(0), 0\} \times \left(\tilde{v}_1(\alpha_1) -  \sum_{j \in \chat^\circ(1)}|\alpha_j(C|M) - \alpha_1(C)|\right)\\
 & \thickapprox 0.9 - 0.6094 >0
 \end{split}
\end{equation*}

Each agent's reaction to the message is indicated in Figure \ref{fig:FULL}: Agent 2, 3, 4, 9, 10 choose to stay in silence, and 5, 6, 7, 8 choose to openly disapprove the veracity of the message.

The example illustrates several insights that the model would provide. First, among the ten agents, only one has a (relatively) strong belief in the veracity of $M$; the agents who deem $M$ a low credence form an absolute majority (agent 2, 3, 5 -- 10). Nevertheless, the message still travels through the network and reaches every chatroom. Here, the behavior of agent 2 (and agent 3, who is a ``replica'' of agent 2) deserves some exploration. Agent 2 does not quite believe the credence of $M$, indeed, $\theta_2 = \alpha_2(M)=0.26$. Yet in the chatroom that he receives the message, he chooses to be silent because there are two agents in the chatroom (1 and 4) who assign $M$ a higher credence, and his sensitivity of peer pressure makes him stay in silence.\footnote{Note that 0.26 is closer to 0.5; hence if agent 2 does not care about others' opinion (i.e., $\lambda_2 = 0$), he will also choose to be silence. Yet $\lambda_2 = 1$ makes him sensitive to the opinion in the chatroom: if, for example, other agents deem a lower credence of $M$, agent 2 may also choose to openly denounce $M$.} 

It might be even more strange that agent 2 chooses to send the message, given that he hardly believes in both $M$ and $C$. Yet in our model, his choice is rational: agent 2 faces two receivers, 5 and 6, whose belief in $C$ is even weaker. Hence it is to agent 2's interest to send the message to them to assimilate their belief to his own. 

\section{Results}

It is straightforward to see the following result.
\begin{observation}
For generic $p$, when $GB(\mathbf{T}, (\Theta_i, \lambda_i, p^i)_{i \in \mathbf{I}}) \neq \emptyset$, it is a singleton.
\end{observation}







\begin{proposition}\label{prop:disprov}
Given $\mathbf{T}$, $\Theta$, $p$, and $\lambda$, $\lambda^\prime$ such that $p^i$ is a Dirac for each $i$ and  $\lambda^\prime \geq \lambda$. Suppose that the following two conditions are satisfied:
\begin{itemize}
\item Both $GB(\mathbf{T}, (\Theta_i, \lambda_i, p^i)_{i \in \mathbf{I}})$ and $GB(\mathbf{T}, (\Theta_i, \lambda^\prime_i, p^i)_{i \in \mathbf{I}})$ are singleton;

\item For a fixed chatroom $\chat$ and each $i \in \chat^o$, $0$ is not the closest action to $\mathbb{E}_{p^i}(\mathsf{N}(\cdot))$.
\end{itemize}
Then for each non-terminal $i \in \chat^\circ$, if $i$ chooses to send the message in the equilibrium in $GB(\mathbf{T}, (\Theta_i, \lambda_i, p^i)_{i \in \mathbf{I}})$, so she does in  the equilibrium in $GB(\mathbf{T}, (\Theta_i, \lambda^\prime_i, p^i)_{i \in \mathbf{I}})$.

\end{proposition}

This result can be easily generalized: each $p^i$ does not have to be Dirac. Further, the statement is robust: if each $\ell_i$ is large enough, then the term ``\emph{each} $i \in \chat^o$'' could be replaced by ``most (or enough) $i \in \chat^o$''. Yet note that this statement does not hold if for some $i$, $\lambda_i < \lambda_i^*$.

To show this statement, we need the following lemma.
\begin{lemma}\label{lem:0shrink}
Given $b_i =\mathbb{E}_{p^i}(\mathsf{N}(\cdot)))$. Suppose that $0$ is not the closest action to $b_i$, and for some $\lambda_i > 0$, $\theta_i \in \mathbb{A}_i(0;b_i, \lambda_i)$. Then for each $\lambda^\prime_i < \lambda_i$, $\theta_i \in \mathbb{A}_i(0;b_i, \lambda^\prime_i)$.
\end{lemma}
\begin{proof}
Without loss of generality, suppose that 0.5 is the closest action to $b_i$. Then, $\theta_i \in \mathbb{A}_i(0;b_i, \lambda_i)$ implies that
\begin{align}
\theta_i + \lambda_i b_i &\leq  |0.5 - \theta_i| + \lambda_i |0.5 - b_i|  \label{eq:first} \\
\theta_i + \lambda_i b_i &\leq  (1- \theta_i) + \lambda_i (1 - b_i)  \label{eq:second}
\end{align}
Since 0.5 is the closest action to $b_i$, it follows that $b_i \in (0.25, 0.75)$, which implies that $b_i \geq |0.5 - b_i|$. Then (\ref{eq:first}) implies that $\lambda_i \leq \frac{|0.5 - \theta_i|- \theta_i}{b_i - |0.5 - b_i|}$. Since $\lambda_i^\prime < \lambda_I$, it follows that $\lambda^\prime_i < \frac{|0.5 - \theta_i|- \theta_i}{bi - |0.5 - b_i|}$, and, consequently,
\begin{equation*}
\theta_i + \lambda^\prime_i b_i < |0.5 - \theta_i| + \lambda^\prime_i |0.5 - b_i| 
\end{equation*}
For (\ref{eq:second}), note that when $b_i \in [0.5, 0.75)$, it can be proved in a similar manner that $\theta_i + \lambda^\prime_i b_i \leq  (1- \theta_i) + \lambda^\prime_i (1 - b_i)$. Suppose $b_i \in (0.25, 0.5)$. Then $b_i < 1-b_i$. Note that it follows from (\ref{eq:first}) that $\theta_i \leq |0.5 - \theta_i|$, i.e., $\theta_i \in [0, 0.25)$, which implies that $\theta_i < 1- \theta_i$. Hence, for any $\lambda^\prime$, $\theta_i + \lambda^\prime_i b_i \leq  (1- \theta_i) + \lambda^\prime_i (1 - b_i)$. Here we have proved that $\theta_i \in \mathbb{A}_i(0;b_i, \lambda^\prime_i)$.
\end{proof}

Lemma \ref{lem:0shrink} is partially a complementary of Proposition \ref{prop:extension} because it focuses on a specific case, though it is not difficult to generalize the result. The underlying intuition is straightforwar: if an action (here it is $0$) is chosen even though it is not the closest to the average credence in one's belief, it is because one's sensitivity to peer pressure is not high enough to bent her choice; therefore, when the sensitivity is even lower, she will clearly not bent herself either. 

\begin{proof}[Proof of Proposition \ref{prop:disprov}]
Combining Lemma \ref{lem:0shrink} and Proposition \ref{prop:extension}, we can see that if $\lambda^\prime \geq \lambda$, and both $GB(\mathbf{T}, (\Theta_i, \lambda_i, p^i)_{i \in \mathbf{i}})$ and $GB(\mathbf{T}, (\Theta_i, \lambda^\prime_i, p^i)_{i \in \mathbf{I}})$ are non-empty and singleton, then, since for each $i \in \chat^o$, $0$ is not the closest action to $\mathbb{E}_{p^i}(\mathsf{N}(\cdot))$, everyone $i \in \chat^o$ who chooses $s$ in the equilibrium $GB(\mathbf{T}, (\Theta_i, \lambda^\prime_i, p^i)_{i \in \mathbf{I}})$ also chooses $0$ in $GB(\mathbf{T}, (\Theta_i, \lambda^\prime_i, p^i)_{i \in \mathbf{I}})$. In other words, the number of agents in $\chat$ who choose $0$ does not increase. Since, given one's belief fixed, only the number of disapprovals can stop the sending of $M$, it follows that $i \in \chat^\circ$, if $i$ chooses to send the message in the equilibrium in $GB(\mathbf{T}, (\Theta_i, \lambda_i, p^i)_{i \in \mathbf{I}})$, so she does in  the equilibrium in $GB(\mathbf{T}, (\Theta_i, \lambda^\prime_i, p^i)_{i \in \mathbf{I}})$.
\end{proof}

\begin{proposition}
Given $\mathbf{T}$, $\Theta$, $p$, and $\lambda$ such that for a chatroom $\chat(\imath)$ in $\mathbf{T}$ with $\imath$ as its root, $\Theta_i$ is a singleton for each $i \in \chat(\imath)$. Suppose that $GB(\mathbf{T}, (\Theta_i, \lambda_i, p^i)_{i \in \mathbf{I}})$ is a singleton. Now consider $\tilde{\Theta}$ and $\tilde{p}$ satisfying
\begin{itemize}
\item[(i)] $\tilde{\Theta}_i = \Theta_i$ for each $i$ except an agent $i^* \in \chat^\circ(\imath)$, and $\tilde{\Theta}_{i^*}$ is a singleton;

\item[(ii)] $\tilde{p}^i = p^i$ for each $i= i^*$ or $i \not\in \chat(\imath)$; 

\item[(iii)] For each $i \in \chat(\imath)\setminus \{i^*\}$, $\marg_{\Theta_{j \neq i, j \neq i^*}}\tilde{p}^i= \marg_{\Theta_{j \neq i, j \neq i^*}}p^i$
\end{itemize}
Let $\Theta_{i^*} = \{\theta_{i^*}\}$ and $\tilde{\Theta}_{i^*} = \{\tilde{\theta}_{i^*}\}$. Suppose that $GB(\mathbf{T}, (\tilde{\Theta}_i, \lambda_i, \tilde{p}^i)_{i \in \mathbf{I}})$ is a singleton. If $\imath$ sends $M$ in the equilibrium in $GB(\mathbf{T}, (\Theta_i, \lambda_i, p^i)_{i \in \mathbf{I}})$ yet she chooses not to send it in the equilibrium in $GB(\mathbf{T}, (\tilde{\Theta}_i, \lambda_i, \tilde{p}^i)_{i \in \mathbf{I}})$, then $\tilde{\theta}_{i^*}$ and $\theta_{i^*}$ satisfy the following condition: there are $\overline{\theta}$ and $\underline{\theta}$ with $\underline{\theta} > \overline{\theta}$ such that
\begin{itemize}
\item[(1)] When $\theta_{i^*}, \tilde{\theta}_{i^*} <\underline{\theta}$ or $\theta_{i^*}, \tilde{\theta}_{i^*} > \overline{\theta}$, $\tilde{\theta}_{i^*} <\theta_{i^*}$;

\item[(2)] When $\tilde{\theta}_{i^*}, \theta_{i^*} \in (\underline{\theta}, \overline{\theta})$,  $\tilde{\theta}_{i^*} > \theta_{i^*}$.
\end{itemize}

\end{proposition}

\begin{proof}
The three conditions imply that, if $\imath$ chooses not to send $M$ under $\tilde{\Theta}$ and $\tilde{p}$, the only reason is that by sending it, the posterior distance from the beliefs of agents in $\chat^\circ(\imath)$ to her own would be larger. To see when it becomes larger, we need to analyze when the following function decreases:
\begin{equation*}
\nu(\theta) = \left|\alpha_\imath(C) - \frac{\theta - \mu_{M|\urcorner C}}{\mu_{M|C}- \mu_{M|\urcorner C}}\right| - \left|\alpha_\imath(C) - \frac{\mu_{M|C}}{\theta}\times \frac{\theta - \mu_{M|\urcorner C}}{\mu_{M|C}- \mu_{M|\urcorner C}}\right|
\end{equation*}
Note that here, $\theta$ represents agent $i^*$'s credence of $M$. To simplify the symbols, in this proof we use $x$ to denote $\theta$, and we let $\alpha_\imath(C) = \tau$, $\mu_{M|C} = \alpha$, and $ \mu_{M|\urcorner C} = \beta$. Then 
\begin{equation}
\nu(x) = \left|\tau - \frac{x-\beta}{\alpha - \beta}\right| - \left|\tau - \frac{\alpha(x - \beta)}{x(\alpha - \beta)}\right|
\end{equation}
First, one can see that $\nu(\theta)$ is a continuous piecewise function: the interval $(\beta, \alpha)$ is divided into three sub-intervals and the function on each interval is:
\begin{equation*}
\nu(x) =
\begin{cases}
\frac{(\alpha- x)(x-\beta)}{x(\alpha-\beta)} & \text{when $x \leq \frac{\alpha\beta}{\alpha - \tau(\alpha - \beta)}$}\\
2\tau - \frac{(\alpha+ x)(x-\beta)}{x(\alpha-\beta)} & \text{when $x \in (\frac{\alpha\beta}{\alpha - \tau(\alpha - \beta)}, \beta + \tau(\alpha - \beta)]$} \\
\frac{(x- \alpha)(x-\beta)}{x(\alpha-\beta)} & \text{when $x >  \beta + \tau(\alpha - \beta)$}
\end{cases}
\end{equation*}

On the first interval, since $\frac{d\nu(x)}{dx} =\frac {\alpha\beta - x^2}{x^2(\alpha - \beta)}$, one can see that if $\sqrt{\alpha\beta} \geq \frac{\alpha\beta}{\alpha - \tau(\alpha - \beta)}$, $\nu(x)$ increases on that interval; otherwise $\nu(x)$ until $x = \sqrt{\alpha\beta}$ and then decreases. On the second interval, since  $\frac{d\nu(x)}{dx} =- \frac {\alpha\beta + x^2}{x^2(\alpha - \beta)} <0$, it always decreases. On the third interval, since $\frac{d\nu(x)}{dx} =\frac{x^2- \alpha\beta}{x^2(\alpha - \beta)}$, similar to the first interval, one can see that if $\sqrt{\alpha\beta} \leq \beta + \tau(\alpha - \beta)$, $\nu(x)$ increases; otherwise it decreases until $x = \sqrt{\alpha\beta}$ and then increases.

To summarize, by letting
\begin{align*}
\underline{\theta} = \min \left\{\sqrt{\alpha\beta}, \frac{\alpha\beta}{\alpha - \tau(\alpha - \beta)}\right\}\\
\overline{\theta} = \max \left\{\sqrt{\alpha\beta}, \beta + \tau(\alpha - \beta)\right\}
\end{align*}
one can see that $\nu(x)$ increases on $(\beta, \underline{\theta}]$ and $(\overline{\theta}, \alpha)$, and decreases on $[\underline{\theta}, \overline{\theta}]$. The configuration of $\nu$ is sketched in Figure \ref{fig:GAP}.

\begin{figure}[h!] 
\centering
  \includegraphics[width=0.6\columnwidth]{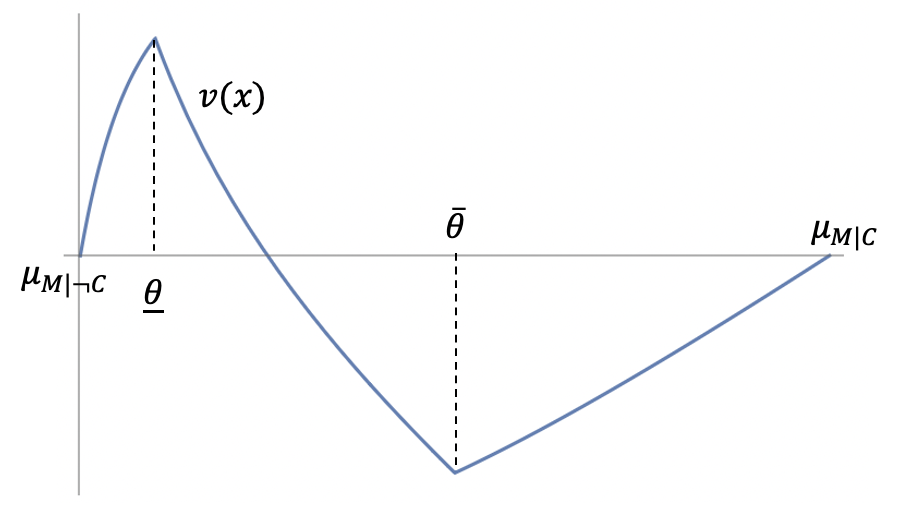}
  \caption{The sketch of $\nu(\theta)$}
  \label{fig:GAP}
\end{figure}
Here, we have shown the statement. Indeed, when $\theta_{i^*} < \underline{\theta}$ or $\theta_{i^*}, \tilde{\theta}_{i^*} < \overline{\theta}$, decreasing $i^*$'s credence of $M$ would enlarge the gap, while when $\theta_{i^*}, \tilde{\theta}_{i^*} \in (\underline{\theta}, \overline{\theta})$, increasing $i^*$'s credence of $M$ would enlarge the gap.
\end{proof}


\section{Discussion}

\subsection{Two measures of similarity}


Note that the two roles employ different measures of similarity. In the receiver’s role, similarity is defined by the distance between an agent’s belief and the \emph{average} belief, whereas in the sender’s role, it is measured by the sum of the deviations between the agent’s belief and those of all other agents. A natural question, then, is whether a unified measure could be used. Below, we explain why we choose not to adopt such a formulation.

The unified decision principle of this paper is that agents seek to maximize belief alignment with their chatrooms. Yet in the two roles the term ``alignment'' has different nuances. In the receiver’s role, we emphasize the desirability of being moderate and of avoiding picking side, as observed in empirical research \citep[see, for example,][]{sea51}, which is faithfully captured in (\ref{puti}). 

We may instead define the psychological utility as
\begin{equation}\label{putialter}
\tilde{u}_i (a, \alpha) = -\left(|a_i - \alpha_i(M)| + \frac{\lambda_i}{\#\chat_R(i)}\sum_{j\in \chat_R\circ(i)}\left|a_i -  \alpha_j(M)\right|\right)
\end{equation}
Adopting a utility function as in  (\ref{putialter}) gives a similar pattern as in Proposition \ref{re2}. Yet in some cases, the two utility functions recommend different optimal actions. For example, consider a chatroom constituted by three agents, $1$, $2$, and $3$, where $\alpha_1(M) = 0.2$, $\alpha_2(M) = 0$ and $\alpha_3(M) = 1$. Suppose that $\lambda_1 = 1$. By adopting the utility function as in (\ref{puti}), silence is the best response for agent $1$, since $u_1(\alpha, 0) = -0.7$, $u_1(\alpha, 0.5) = -0.3$, and $u_1(\alpha, 1) = -1.3$. Yet if we adopt the utility function in (\ref{putialter}), the best response is to disapprove it openly, since $\tilde{u}_1(\alpha, 0) = -0.7$, $\tilde{u}_1(\alpha, 0.5) = -0.8$, and $\tilde{u}_1(\alpha, 1) = -1.3$.

Upon reflection, one might notice that both selections make sense. Our model chooses silence since the similarity measure in (\ref{puti}) emphasizes moderation, which is closer to the center, while  (\ref{putialter}) emphasizes on inclining to the opinion held by the majority, which is disapproval. In the reality, if the chatroom is so polarized, an agent in a chatroom formed by acquaintances with long-term off-line relationship might be more hesitant to clearly pick side and place importance on maintaining balance. Therefore, we adopt the formula in (\ref{puti}) as the psychological utility function.


For the sender's role, we want to capture the desirability for the sender to assimilate others' beliefs to hers. Here, the emphasize has to be person-wise. This purpose could be missed if we use the gap between one's own belief and the average one since some singularity might be caused by the ``mutual cancelling''. For instance, consider $\chat_S(1) = \{1,2,3\}$, that is, agent 1 is the root and 2 and 3 are her immediate successors, $\alpha_1(C) = 0.5$, and the beliefs of $2$ and $3$ depicted in Table \ref{TAB06}. \footnote{Here, our point is to illustrate the different recommendations of the two distances. Hence we ignore the restrictions such as the consensus on evidence relationshiops.}

\begin{table}[ht!]
\centering
 \begin{tabular}{c c c}
\hline
    & $C$  &  $C|M$\\ 
 \hline
$\alpha_2$ & $0.4$  & $0.44$\\ 

$\alpha_3$ & $0.6$  & $0.63$\\

 \hline
\end{tabular}
\caption{Beliefs of agents in $\chat(1)$}
\label{TAB06}
\end{table}
One can see that by sending the message, 2's worldview is closer to 1's; 3's worldview is farther, yet the difference is smaller than the improvement of 2's. Therefore, sending $M$ is better than not sending. Yet if we compute the distance between $1$'s belief and the average of others' beliefs, one can see the payoff for not sending is $0$ and sending is $-0.035$, suggesting that not sending is worse that sending, which is contradictory to the intuition. 

\subsection{More flexible network structure}
\begin{figure}[htbp]
  \centering
  \tikzset{
    every node/.style={font=\large},
    edge from parent/.style={draw, -{Stealth[scale=1.0]}, semithick, shorten >=1pt},
    level 1/.style={sibling distance=40mm, level distance=15mm},
    level 2/.style={sibling distance=20mm, level distance=15mm},
    level 3/.style={sibling distance=15mm, level distance=15mm}
  }

  \begin{tikzpicture}[grow=down]
    \node {5}
      child {node {6}}
      child {node {2}
        child {node {1}}
        child {node {3}
          child {node {7}}
          child {node {8}}
        }
        child {node {4}
          child {node {9}}
          child {node {10}}
        }
      };
  \end{tikzpicture}

  \caption{Directed tree rooted at 5.}
  \label{fig:tree5}
\end{figure}
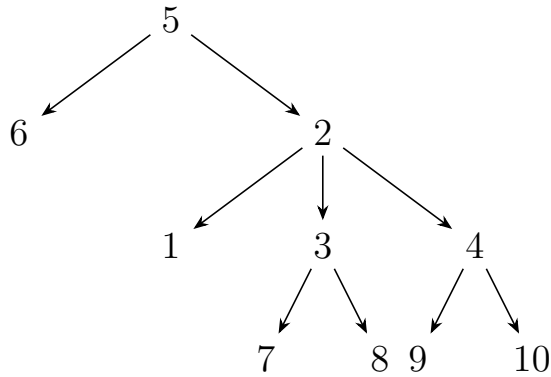

In the example in Section \ref{sec:eq-ex}, one may observe that the most ``subversive'' agents, namely agents 5 -- 10, are among the final ones to receive $M$. Their credence of $M$ is the lowest among $\mathbf{I}$. Further, since they are grouped together, they can openly disapprove the message. Therefore, if the message reaches them first, for example, $5$ is the root, the tree would be like in Figure \ref{fig:tree5}, and the message will not be sent at all. 

Hence, from the viewpoint of an outsider information manipulator, if its purpose is to make $M$ reach as many people as possible, it should (1) target the one who has a relatively high credence of $M$, (2) make sure that the disbelievers appear latter in the network, (3) try to ``dilute'' the disbelievers, for example, to mix true believers with disbelievers in each chatroom and never let the disbelievers dominate.

This observation is relevant, but cannot be discussed in our present model. Yet we can extend our framework to facilitate such discussion. Instead of a given ordered tree, we can start from a connected, undirected graph $G=(I, E)$, where $I$ is the set of individuals and $E$ the set of edges.\footnote{Here, by \emph{connected} we mean that every pair of agents are connected by a path.} For each $i \in I$, let 
$E(i):=\{j \in I: \{i, j\} \in E\}$ denote the set of \emph{neighbors} of $i$. We assume that $G$ is irreflexive, i.e., no agent is her own neighbor. We restrict attention to graphs that satisfy the structural conditions: for each $i, j, j^\prime \in I$ with $\{i, j\}$, $\{i, j^\prime\} \in E$,
\begin{itemize}
\item[\textbf{T1}.] $\{j, j^\prime\} \in E$, and

\item[\textbf{T2}.] For each $k \in I$ with $\{i,k\} \not\in E$, if $k$ is connected to one in $j$ and $j^\prime$, it is not connected to the other; formally, $\{k,j\} \in E \Rightarrow \{k,j^\prime\} \not\in E$.
\end{itemize}

\begin{figure}[h!] 
\centering
  \includegraphics[width=1\columnwidth]{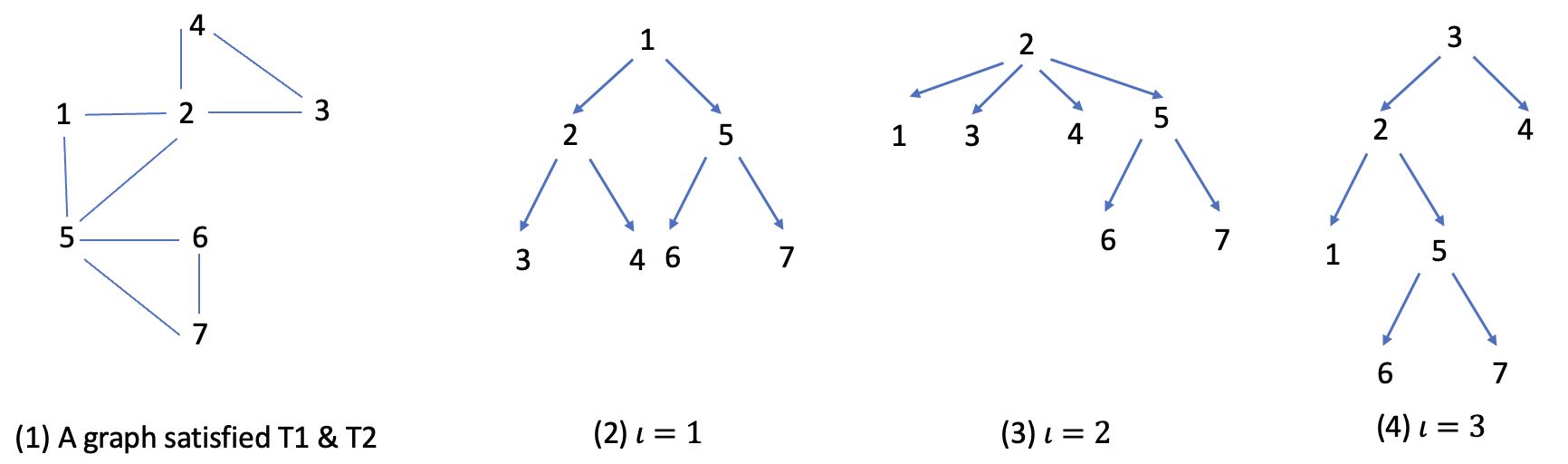}
  \caption{From a undirected graph to a tree with by assigning a root}
  \label{fig:UN00}
\end{figure}

T1 and T2 ensure that the network can be consistently transformed into an ordered tree once a root is specified. To be specific, given $\imath$ as the root, $E(\imath)$ is the set of $i^o$'s immediate successors, and for each $i \in E(\imath)$, its immediate successor is $E(i) \setminus E(\imath)$, etc. The generated tree is denoted as $\mathbf{T}(\imath)$. As an illustration, consider the graph in Figure \ref{fig:UN00} (1). The graph is connected and satisfies T1 and T2. By assigning an agent to be the root, the graph always generates an ordered tree: Figure \ref{fig:UN00} (2) -- (4) depict the trees generated by assigning agents $1$, $2$, and $3$ as the root, respectively.

In this manner, we could describe the flow of information on a network of chatrooms from any initial sender $\imath$(i.e., root), and we can discuss the consequence by targeting a specific as the root.

\end{document}